# Toward a formalization of artifacts in GFO

## *PREPRINT VERSION*

Hanna Fiegenbaum[a]

[a]*Leipzig University, Institute for Medical Informatics, Statistics and Epidemiology, Härtelstr. 16-18, 04107 Leipzig, Germany*
*E-mail: hanna.fiegenbaum@uni-leipzig.de*
*ORCiD ID: Hanna Fiegenbaum 0000-0002-2462-5924*

**Abstract**. The General Formal Ontology (GFO) is a top-level ontology that is designed to formally describe different domains of reality. Most recent advancements within GFO have been made in defining its modules of space and material objects, defining its functions, and a module for integrated data semantics. In this paper, I further develop the GFO towards the integration of artifacts, which are material objects that are intentionally made for a certain purpose. I discuss recent advancements in artifact ontology in philosophy and formal ontology alike, and provide basic categories and axioms for artifact description in GFO, while considering existing work within its space module, object-process integration and function module.

Keywords. Artifact Ontology, GFO, Material Object, Formalization

## 1. Introduction

Artifacts are commonly understood as objects intentionally made, usually with specific purposes in mind (Hilpinen 1992, Preston 2009). This category encompasses a wide spectrum of items including tools like hammers, means of transportation such as boats and cars, technological entities like computers, software or phones, personal care products like nail polish, printed products like books, the built environment, stationary goods such as pencils, and toys and clothes and many more.

Beth Preston (2022) references Risto Hilpinen's definition of artifacts: "An artifact may be defined as an object that has been intentionally made or produced for a certain purpose. (Hilpinen, 1992, 2011)". In her publication "The Ontology of Artifacts" (2004), Lynne Baker echoes that definition when asserting that artifacts are "objects intentionally made to serve a given purpose. " (Baker, 2004:1).

This standard definition of artifacts emphasizes two distinctive qualities: firstly, their intentional creation, and secondly, their capacity to be employed purposefully. Recent studies by Juvshik (2021a, 2021b, 2022) have shown how the dependence on intentionality in defining artifacts can be further clarified. Furthermore, work by Juvshik (2021b) and others questions whether the presence of functional attributes necessarily constitutes an inherent characteristic of artifact type descriptions. However, others (Bahr, 2019; Mizoguchi, 2009) have developed classification systems for artifacts according to their functions. And for industry products and technical artifacts, a specific subset of artifacts, functional attributes of artifact types are commonly included in their descriptions (Burek, 2021; Mizoguchi, 2009).

In formal ontology, several approaches exist to formally specify artifacts and their features for different top-level ontologies. The focus has been on defining engineered objects and their classification for manufacturing environments. One prominent example is the Descriptive Ontology for Linguistic and Cognitive Engineering, DOLCE, for which artifact descriptions have been developed by Borgo and Vieu (2009), Borgo (2008, 2014) and others. Another top-level ontology, Basic Formal Ontology, describes "engineered units" which are manufactured artifacts (Smith, 2012). Weigand (2021) starts developing an ontology for artifacts conceived for integration into Design Science Research within the top-level ontology UFO. YAMATO, another top-level ontology developed in a common effort by philosophers and engineers alike, focuses on formalizing function descriptions for artifacts as roles (Mizoguchi et al., 2022). For the top-level ontology GFO, in its most recent version 2.0, Burek et al (2021) provide a formalization of functions of objects, where function bearers can also be artifacts. However, a formal description of artifacts for GFO has not yet been suggested.

The present paper aims to fill this gap. The objective of this paper is to develop GFO 2.0 further towards the integration of material objects that are intentionally made by intentional agents for a certain purpose, in other words, of artifacts. This requires integrating existing modules from Loebe et al. (2021) on formalizing material objects as well as Burek (2007) and Burek et al. (2021) who have developed an account of functions, encompassing artifact functions equally, and Herre (2019), who provides an account of properties, property values and attributives. To develop an account of artifacts building on existing modules, I proceed as follows: At first, I will lay out philosophical considerations of artifacts as material objects while showing that GFO already satisfies the integration of aspects necessary for describing material objects and their material composition. Further, by drawing on recent advances in the metaphysics of artifacts by Juvshik (2021, 2022) and others, I will develop basic artifact categories and axioms by building on the foundation of GFO existing modules. Lastly, I will consider further developments necessary to describe elaborate designing and planning processes of artifact production.

## 2. GFO

The General Formal Ontology (GFO) is a formal-ontological framework that aims to provide a systematic, comprehensive foundation for modeling and representing knowledge across various fields, including biology, medicine, engineering, and social sciences, among others. GFO 2.0 is its most recent version. GFO was mostly developed from research at the University of Leipzig, extending to other academic institutions and actors all around the world developing it further[1]. GFO integrates different levels of reality and types of entities, including objects, processes, and relationships, within a coherent structure. It distinguishes between different categories such as physical, abstract, and social entities, providing a basis for detailed representation of complex systems and phenomena. GFO is one among a few top-level formal ontologies, such as BFO[2], UFO[3], DOLCE[4] and YAMATO[5]. A top-level ontology, also known as a foundational ontology, provides a high-level and domain-independent framework that defines basic, general categories and relations applicable across various fields and types of ontologies. The purpose of a top-level ontology is to offer a universal structure that can serve as a foundation for developing more specific, domain-specific ontologies, ensuring consistency and interoperability among them. Most recent advancements within GFO have been made in defining its modules of space and material objects (Loebe et al., 2021), defining its functions (Burek et al., 2021) and a module for integrated data semantics (Herre, 2016).

---

[1] https://www.imise.uni-leipzig.de/institut/Projekte/GFO
[2] https://basic-formal-ontology.org/
[3] https://ontouml.readthedocs.io/en/latest/intro/ufo.html
[4] https://www.loa.istc.cnr.it/dolce/overview.html
[5] https://www.hozo.jp/onto_library/upperOnto.htm

## 3. Artifacts as material objects

Artifacts are material objects made by intentional agents with an intention to build an artifact of a specific artifact kind. Their being material objects on the one hand, and intentionally made objects embodying human knowledge and capacities on the other, has brought them the characterization of having a "dual nature" (Meijers & Houkes, 2006): Artifacts can be characterized as material objects, occupying a position in time and space, having boundaries and formal, structural and material characteristics as any other material object. But they are also tangible results of the intentional actions to build a thing of a specific kind, with specific features or capacities. These two sides, of being enduring material objects, yet intentionally made and bearers of human knowledge, have been said to constitute their "dual nature". Thus, what differentiates artifact kinds from natural kinds, is that they are intentionally made and have makers who have intentions. What differentiates artifact kinds from institutional and social kinds, in contrast, is that they persist in time and space without human beings having to repetitively enact them as may be the case for social or institutional kinds. Hence, it is important to notice that not all human-made objects are artifacts. Social and institutional kinds such as the law or a societal role are human-made and, depending on the kind, they can be intentionally made, but they are not tangible objects like artifacts. Actions are intentional as well, but not tangible results like artifacts either. Artworks, even though they are human-made and intentionally made, seem to behave differently than artifact kinds and more like social and institutional kinds, and deserve a separate definition. Works of art or institutions such as laws and regulations which are undoubtedly intentionally made as well, can come in different forms and modes of being than artifacts. Therefore, I would like to exclude artworks from the following specification of artifacts, because they deserve a separate discussion. In what follows, I would like to focus on the class of material objects that persist through time and are intentionally made to fulfill a certain role or purpose, which are commonly termed artifacts. Because artifacts are material objects, I will start by recalling the existing definitions for material objects in GFO in order to build a description for artifacts based on that.

## 4. Material objects

The concept of a material object has been prominently shaped in the philosophical and ontological debate by Peter Van Inwagen in his book "Material Beings" (1990). In his work, Van Inwagen addresses the question of what exactly material objects are and what can be considered a material object. His approach involves committing early on to a description of what qualifies as a material object and then examining various cases to see if they meet the criteria established at the beginning. He provides the following description and sufficient conditions for material objects:

"A thing is a material object if it occupies space and endures through time and can move about in space (literally move about, unlike a shadow or a wave or a reflection) and has a surface and has a mass and is made of a certain stuff or stuffs" (Van Inwagen, 1990:17).

According to Van Inwagen, an object is a material object if it occupies space, persists through time, can move in space, has a surface and mass, and is made of a certain "stuff." Van Inwagen further specifies material things as entities that endure through time and typically undergo change over time (Van Inwagen, 1990:4). While Van Inwagen considers mostly living beings as those material beings that are relevant for the ontological realist, many of the features he introduces in his definition of material objects have been echoed in the debate by others and are also reflected in formal-ontological definitions of material objects. Material objects persist in time and space and occupy a certain position in space. They have boundaries, a surface, a mass, density and consist of material and possibly spatial parts. All of these features can be found in Loebe et al. (2021) who introduce descriptions for the definition of material objects in GFO 2.0.

## 5. Material objects in GFO

The General Formal Ontology lists properties and relations to describe material objects relative to space in its Space Module (Loebe et al., 2021) . A material object exists in space and time, consists of stuff, has mass, density, a lifespan, and can possess parts. The following is an excerpt of relations defined to describe material entities and then further material objects by Loebe et al. (2021):

"Fluid(x) (x is a fluid entity)
Gas(x) (x is a gaseous entity)
MatE(x) (x is a material entity)
ML(x) (x is a material line)
MOb(x) (x is a solid material entity with smooth boundary)

MS(x) (x is a material surface)
MStr(x) (x is a material structure)
MVert(x) (x is a material vertex/point)
ObSit(x) (x is an object-situation)
Stuff(x) (x is an amount of material stuff)

consists_of(x,y) (material object x consists of the stuff y)
contained_in(x,y) (material object x is contained in material object y)
environ(x,y) (x is an environment of y)
has_density(x,y) (x has a density quality y)
has_mass(x,y) (x has a mass quality y)
lifetime(x,y) (x is the life time of the material object y)
maxbd(x,y) (x is a maximal material boundary of y)
mbd(x,y) (x is a material boundary of y).
mpart(x,y) (x is a material part of y)
natmbd(x,y) (x is a natural material boundary of y).
occ(x,y) (material object x occupies space region y)
occbd(x,y) (material boundary x occupies spatial boundary y)
touch(x,y) (material boundaries x and y touch (or are in contact with) each other)"

MatE(x) describes a material entity x that can be either a solid object, a fluid, or a gaseous entity. All material entities consist of stuff, and have mass and density. Loebe et al. (2021) use the term "material object" for solid material entities. A material object MOb(x) is a solid material entity with a smooth boundary that occupies a specific spatial region.

"(2) General axioms on material entities.

M1.     $\forall x\ (MOb(x) \lor \quad\quad Fluid(x) \lor Gas(x) \rightarrow MatE(x))$

M2.     $\forall$ x (MatE(x) → $\exists$ y (Stuff(y) $\land$ consists_of(x,y)))
M3.     $\forall$ x (MatE(x) → $\exists$ yz (has_mass(x,y) $\land$ has_density(x,z)))"

The authors further provide basic axioms for material objects and their relation to space. The predicate SReg(x) defines a space region that a material object occupies and Conn(x) the property of its connectedness (see M4). An object situation ObSit(y) describes the surrounding environment of a material object. This can be a forest for an animal or a river for a fish, containing other material entities of various kinds - fluid like water, gaseous like air, or material objects such as plants (see M9). A material object can be defined in a spatial environment by matching its boundaries with boundaries within a space region (see M7).

"M4.     $\forall$ x (MOb(x) → $\exists$ y (SReg(y) $\land$ occ(x, y) $\land$ Conn(y))) Every material object occupies a connected space region. A stronger notion of connectedness, called material connectedness, requires forces that hold the material object together.

M5.     $\forall$ x (MOb(x) → $\exists$ y (mbd(y,x))) Note that the boundary that must exist is not assumed to be maximal.

M6.     $\forall$ xyz (MOb(x) $\land$ mbd(y,x) $\land$ mpart(z,y) → mbd(z,x)) Every material part of the boundary of a material object is itself a material boundary.

M7.     $\forall$ xyzu (MOb(x) $\land$ occ(x,y) $\land$ mbd(z,x) → $\exists$ !u       (sb(u,y) $\land$ occ(z,u))) If a material object occupies a space region, then any boundary of the material object occupies a uniquely determined boundary of the occupied space region.

M8.     $\forall$ x (MOb(x) → $\exists$ y (environ(y,x))) For every material object there exists an environment.

M9.     $\forall$ xy (MOb(x) $\land$ environ(y,x) → ObSit(y) $\land$ contained_in(x,y)) The environment of a material object is an object-situation that contains this material object. This object-situation is not uniquely determined and can be arbitrarily extended. An object-situation can be understood as a complex material entity that may contain material entities of different states of aggregation (solid, gasses, fluids). The environment of a fish, for example, may contain a part of a river with the water, water plants and stones at the river's ground."

Further, as follows from their being defined by boundaries in an environment, material objects do not share parts with their environment:
"M15.   $\forall$ xy (MOb(x) $\land$ environ(y,x) → ¬ $\exists$ z (MatE(z) $\land$ mpart(z,x) $\land$ mpart(z,y))) A material object has no common material part with an environment." (Loebe et al., 2021)

Additionally, it should be clarified that material parts of a material object sit within its occupied space region:
"M13.   $\forall$ xy (mpart (y,x) $\land$ occ(x,z) → $\exists$ u (spart(u,z) $\land$ occ(y,u))) Any material part of a material object occupies a spatial part of the space region occupied by the material object. We do not admit an axiom saying that for every spatial part of the occupied space region there exists a material part that occupies exactly this spatial part." (Loebe et al., 2021).

The description of material objects by Loebe et al. (2021) satisfies the provision of relations necessary to describe these objects relative to their position in space. It allows for these objects to have boundaries, a surface, a mass, density and to consist of material. What is lacking is the integration of material objects with time. It is important to note, that material objects in GFO are considered to be endurants or continuants, that is objects that persist in time and space (Herre, 2019).

## 6.   Endurants and perdurants

Most formal ontologies start with distinguishing endurants and perdurants. Endurants are entities that exist wholly at each moment of their existence, but extend in space. One can think of an endurant as of a material object, such as a rock or a book. At any given moment, the entire book is present; it doesn't exist partially at one moment and then partially at another. In contrast, perdurants are entities that are extended through time by having temporal parts at different moments. They are often thought of as a series of temporal "slices" or parts that together constitute the whole entity across time. We can consider a movie as a perdurant because it unfolds through time and at different times it exists as a different part or slice of itself. Theories of material constitution (Baker, 2004, 2007) seek to explain how material entities, which consist of slices or parts delineated both in space and/or time, retain their identity across these dimensions. Despite undergoing temporal and/or spatial changes in parts, these entities continue to be the same. The relation of material constitution can indeed apply to the segmentation of material entities across both the dimensions of time and space. This attribute makes the concept especially appealing for a theory of artifacts whose ambition is, among other, to account for the replacement of certain parts of a material object while still claiming that the artifact remains the same over time. Let's say a chair changes a leg after its one leg broke, how can we claim it remains the same chair despite its changes in parts?

## 7.    Material Constitution

A debate which tries to solve exactly this puzzle around the question of enduring material objects and changes within their material composition in time and space, is the debate around material constitution (Baker, 2004, 2007). In the debate about material constitution, the central question generally revolves around the identity and persistence of composite material objects.

Initially, there are no strict limits to what qualifies as a composite material object. It could be an artifact like a ship, a living being, a stone, or even a mass of clay or sand. Living beings undergo changes in their composition over time. An animal grows, a plant decays and decomposes. Artifacts can undergo changes in their composition, too. A chair is assembled and disassembled. Therefore the question arises: Does an object remain identical to itself if its form and/or composition change over time, or does it become a different object as a result? Do two objects coexist at the same time and place, and if so, how is this possible, and what is the relationship between these two objects?

A prominent representative in the debate on the material constitution of material objects, particularly artifacts, is Lynne Baker (2004, 2007). To explain the composition of a material object by other material objects, Baker employs the aforementioned relation of material constitution. Through the constitution relation in Baker's ontology, material objects of a new kind, say K, can be brought forth by objects of other kinds, say S, without K being reducible to entities of lower-level kinds such as S. The K, let's say an instance a of the kind chair K, is not identical to but materially constituted by its parts S such as its legs. Hence, changing one of its parts, changes the material base S which constitutes the chair a, but it does not turn the instance a of kind chair K into another object. The instance a of K remains identical, while its parts do not.

Material constitution exists, according to Baker, between instances of different primary kinds: "constitution is a relation between things of different primary kinds" (Baker, 2004:2). According to Baker, this also means that the chair-parts not only constitute the chair but also bring into existence an object of a higher-level primary kind: "constitution brings into existence new objects of higher-level primary kinds than what was there before" (Baker, 2004:2). The chairs legs, seat, armrests, and backrest, stand in relation to the chair in that they bring forth or bring into existence an instance of a higher-level primary kind, the chair. Constitution, at least in Baker's ontology, seems to be a relation that can generate hierarchies between objects. Instances of primary kinds can generate instances of higher-level primary kinds by materially constituting them.

Material constitution teaches us that a material object and the materials that constitute it are not identical. For this very same reason, a change in the constituting material, does not change the identity of the constituted object. The chair's leg that is broken and then replaced by another leg, does not change the identity of the chair as a whole. What, then, accounts for the identity of a chair to be of a specific artifact kind, the chair? According to Baker (2004), this identity of the primary kind chair is defined by the object being intentionally built to be able to perform the "proper function" of that kind, in this case of a chair. That is, if the chair was intentionally built to perform its function, of someone being able to sit on it, it will remain a chair. While there are other identity criteria for artifact kinds available than Baker presents, an identity criterion among all of them is that an object conforms with properties that are defining an artifact kind, be those functional or not.

Most authors (Baker, 2004; Preston, 2009; Houkes and Vermaas, 2010) agree that functions are part of artifact kind definitions. However, others (Juvshik, 2021; Thomasson, 2007) contend that the main characteristic of artifacts is their being results of intentional actions to build an instance of a specific artifact kind and that kind definitions do not necessarily have to contain functional attributes.

I will give the definition of an artifact kind a closer look in the next section. Here it is important to stress the following: While the instance x of the higher-level primary kind "chair" will remain identical to itself, as long as x satisfies the primary kind chair's identity criteria, the chair instance x is not identical to the lower-level primary kinds of its material base but it is materially constituted by them. The instance of the primary kind chair x is not identical to and can not be reduced to the sum of instances of other primary kinds - legs, arms, wood and so on - in its material base.

Baker argues further that the relation of material constitution displays different characteristics than the identity relation. For the identity relation, it holds that identity is reflexive, symmetric, and transitive (If a=b, then a=a (reflexive). If a=b, then b=a (symmetric). If a=b and b=c, then a=c (transitive)). In contrast, Baker (2004) discusses that material constitution is not reflexive and also not symmetric. Suppose a set of bricks constitutes a house. The house, in turn, does not constitute the bricks, hence, material constitution is not symmetric. The house does not constitute itself either or does not solely constitute itself; something else, the bricks, is necessary for that. Therefore, material constitution is not reflexive. Baker does not explicitly discuss transitivity. One could assume that material constitution is at least sometimes transitive. If the house is constituted by the stones, and the house together with other houses constitutes a village, then one could argue that the stones also constitute the village to a certain degree.

To describe the relation between the constituting material and the constituted object, the top-level ontology GFO recognizes a similar relation: "x consists of y" where a material object x is constituted by the material stuff y. If a pile of bricks Y constitutes a house x, then according to GFO, this would be formalized as: x consists of y.

consists_of(x,y) (material object x consists of the stuff y)

Another relation in GFO 2.0 which is of relevance with regard to distinguishing material constitution from an object containing another object is the relation of containment: contained_in(x,y) (material object x is contained in material object y). If the house contains chairs and a table, this relation can be described using the "contained_in(x,y)" relation.

## 8. Object-Process Integration in GFO

While Loebe et al. (2021) describe material objects in GFO relative to space, Herre (2013) addresses the question of their persistence through time. Material objects, including artifact objects, are continuants in GFO (Herre, 2013; 2019).

According to Herre (2013), an enduring object, or a continuant, "in GFO is additionally underpinned by a process". That is, for any enduring object, a process is defined to describe the features of its existence relative to time. The integration axiom of GFO states accordingly that "for every continuant C there exists a process P, the boundaries of which coincide with the presentials, exhibited by C". "Presentials" are "snapshots" or parts of the continuant's existence in time. Herre (2013) cites a cat crossing the street or a cat losing parts. The cat remains the same although it may lose a part. These time-parts of the object (for instance the cat) combine into the object-complementary process P of the continuant's, the object's, existence. A continuant exhibits presentials at certain points in time. These presentials (time parts or snapshots) combine into a process of a continuant's existence. Herre (2013) thus defines an axiom for such an object-process integration:

"For every material continuant C there exists a process Proc( C) such that the process boundaries of Proc(C) coincide with the presentials, exhibited by C, formally,
$\forall$ C (MatCont(C) → $\exists$ P (Proc(P) ∧ lft( C) = tempext(P) ∧ $\forall$ t M (exhib(C,t,M)  procbd(C,t,M)))".

The relations "exhib(C,t,M" and "procbd(P,t,N)" are defined in the following way:

"exhib(C,t,M): the continuant exhibits the presential M at time point t
procbd(P,t,N); N is the process boundary of the process P at the time-point t"

Within the top-level ontology GFO, the processes are the most fundamental category of spatio-temporal individuals, whereas objects and their snapshots (presentials) depend on processes. Because artifact objects are material continuants in GFO, they have processes defined to describe their existence in the same way as any other material continuants. The boundaries of this process coincide with the presentials exhibited by the artifact object. This process description is inherited by artifact objects in virtue of their being material objects which are continuants in GFO. A thorough discussion of the specific integration of artifact objects and their time components such as processes in GFO could be the objective of another research work. While, however, artifact objects would be associated with processes in much the same way as all other continuants, it could be interesting to assess the intentional actions necessary to build artifact instances with regard to their temporal components.

## 9. Departing from material objects to artifacts

Departing from the description of material objects within GFO 2.0, what is additionally needed to specify the class of artifact objects within the class of material objects? Artifact kinds can be specified within material objects as those material objects that are intentionally made by makers which intended to build an artifact of a specific artifact kind K. An account of artifacts as material objects that are intentionally made needs to be able to provide descriptions of what it means for a material object to be intentionally made as an instance of a specific artifact kind K. In the following section, I will first discuss and revise the recent account of artifact objects by Tim Juvshik (2021a, 2021b, 2022) while incorporating approaches by Hilpinen (1992), Thomasson (2007) and others. Drawing on their insights, I will then suggest a basic account for describing artifacts formally within GFO 2.0.

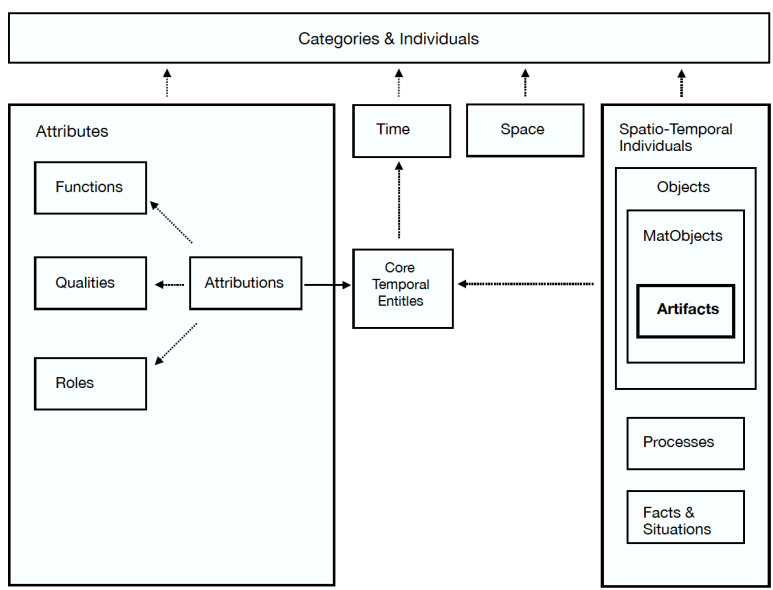

Figure 1. Overview of the main modules of GFO 2.0 and their dependencies.

## 10. Recent advances in the metaphysics of artifacts

Many accounts on artifacts and artifact ontology give necessary and sufficient conditions for that the result of a maker's actions can be called an instance of an artifact kind (Baker, 2004; Houkes & Vermaas, 2009; Thomasson, 2007, 2009; de Ridder, 2007; Hilpinen, 1992; and others). An extensive account is provided in recent work on artifact ontology by Tim Juvshik (2021a, 2021b, 2022). Juvshik (2021a, 2022) gives a comprehensive account of what the dependence of artifacts on human intentions and actions implies. In his 2021 paper titled "Artifacts and Mind-Dependence," Juvshik presents a concept he terms "IDA," which stands for the "intention-dependence of artifacts." The intention-dependence of artifacts states that "something is an artifact of kind K only if it is the successful product of an intention to make an artifact of kind K (2021:1)." Juvshik (2021) therewith defines the particular mind-dependence of artifacts as intention-dependence. But the intention-dependence of artifacts is only one necessary implication of artifact objects. For an object to be classified as a member of the K category, it's not sufficient that the creator simply intended to make an artifact of kind K. It also hinges on the creator's ability to effectively instill the K-relevant characteristics into the object, to the point where the creation of the object as a K can be deemed "successful," as articulated by Juvshik. Accordingly, in his dissertation, Juvshik (2022) calls the intention-dependence of artifacts as only one of six different success conditions for a result of a maker's actions to count as an artifact of kind K (Juvshik, 2022:214).

In the ontological debate around artifacts, it goes undisputed, that artifacts are intentionally made. That they are also made as instances of a certain artifact kind K, evolves as an implication of their maker trying to materialize a concept of a certain thing with certain features they want to build. A maker, before setting out to intentionally build something, say a bike, has a concept of what a bike is and how the result of building a bike will have to look like in order to be qualified as a bike.

Juvshik (2022) covers a similar condition as concept-dependence. Concept-dependence of an artifact claims that a maker has to have a concept of the kind of thing they want to build in order to be able to successfully build an object of that kind: A maker S must have a concept of artifact kind K. From concept-dependence of building an artifact, it follows that there are criterial features that distinguish the properties of that specific artifact kind from other artifact kinds. Juvshik calls these features "criterial features" of an artifact kind K. From wanting to build an artifact of kind K, say a bike, it follows that a maker has to bestow the bike-criterial features on that object that they build. The object has to have wheels, handlebars, and a saddle and be able to carry someone on its saddle in order to be classified as a bike. Juvshik covers these conditions as "kind-relevant features" and "bestowal" of these kind-relevant features on an object. In order for the result of their actions really to be a bike, the maker has to be successful in this bestowal process: A maker S must attempt to bestow some of K's kind-relevant features k1, k2, k3, … kn on an object(s) O. And a maker S must successfully bestow the intended kind-relevant features k1, k2, k3… kn on O to some extent. In order to do that, a maker has to select the adequate material and assemble them in the correct way to build an instance of an artifact kind K that they want to build. Besides concept-dependence, Juvshik lists the attempt to build an instance of an artifact kind K and the successful bestowal of K-criterial features on that instance as necessary conditions to build an instance of an artifact kind K. Juvshik includes as a last condition that an artifact instance is built from existing material: "For any artifact A, there is some pre-existing object(s) O from which A originated as the result of the maker S's attempt to make A (Juvshik, 2022:215) ."

Juvshik's conditions are almost comprehensive in explaining and further specifying what is implied when calling some object the "successful product" of an intention to produce an artifact of kind K. They also take into account and further clarify relevant points from the debate around ontology of artifacts that have been iterated by authors numerous times, such as the dependence of artifacts on their maker's concepts, their maker's intentional actions to build a member of a specific artifact kind, and the implication of criterial features of this artifact kind. Other than Lynne Baker, Juvshik does explicitly exclude the instantiation of functions as a necessary condition for artifact kind identification. Rather, for Juvshik, it is sufficient that an artifact displays the relevant kind-criterial features of a specific artifact kind to be classified as a member of that kind, where these criterial features need not be functional attributes necessarily.

## 11. Beyond concept-dependence

A point which Juvshik does not touch upon, but which can be elaborated further, is that the fulfillment of these conditions of intending and attempting to build an artifact of a specific artifact kind enable the evaluation of the tangible result of these actions with regard to the concept underlying the artifact kind and the specific object. Especially with regard to the points of concept-dependence, kind-membership and criterial features, the debate on metaphysics and ontology of artifacts often overlooks that the intention to build an artifact of a specific kind K, described by a sortal description (Hilpinen, 1992) of K with specific kind-relevant features provides sufficient reason for the evaluation of the tangible result of these intentional actions as a member of this kind K. If someone wants to build a bike, the result of his actions can be evaluated according to the criterial features a bike should possess.

Another aspect that Juvshik does not specify further, is that the process of building an artifact might also go beyond a conceptual representation of that artifact kind, but might actually include a design and a specific model of that artifact kind and an extensive planning process. Weiland et al. (2021) underline that "not much has been written yet on the ontological status of artifacts or their structure" from a Design Science research perspective in the research field of formal ontology. The description of the design, modeling and planning processes of artifacts of a certain artifact kind have indeed not been much developed yet in the field of formal ontology of artifacts. This is, however, not the case in the philosophical discourse. Houkes and Vermaas (2002) have outlined a planning theory of artifact design. They use planning theories of action by Bratman (1987), Audi (1991) and others. Artifacts have also been a research subject in the field of extended mind and embodied cognition with regard to their design (see for instance Richard Menary, 2010) as well as from a perspective of design and technology (de Ridder, 2007). However, Weiland et al. (2021) are right that the wider implications of an artifact design and designing process have so far neither been formally described in the philosophical discourse nor in formal ontology.

In the remaining sections, I will develop artifact categories and relations needed to further specify artifacts as a subclass of material objects for the general formal ontology GFO. This is accomplished by building on and further expanding on insights of what it takes to successfully produce an instance of an artifact kind from both philosophical debate and formal ontology.

## 12. Towards an account of artifacts in GFO

Artifacts are instances of artifact kinds. For a specification of the relations of those, it will be helpful to remind what Burek et al. (2021) refer to, that within GFO, "individuals and categories are disjoint and individuals instantiate categories, not being instantiated themselves" (Burek et al., 2021: 1025). For each individual artifact, there will be an artifact kind that is instantiated by that individual. The following propositions, taken from the paper on formulation of function in GFO by Burek et al. (2021:1025), apply:

(1) $\forall x$  Individual(x) $\leftrightarrow \neg$ Category(x) .

(2) $\forall xy$  is-instance-of(x, y) $\rightarrow$ Category(y) .

(3) $\forall x$  Individual(x) $\rightarrow \exists y$ (Category(y) $\wedge$ is-instance-of(x, y)) .

Individuals and Categories are disjoint sets, but for any individual it holds that there is a category of which that individual is an instance of. Herre (2019) writes accordingly on GFO: "a property is specified by three components (ECat, P, R), where ECat is a category of entities, P is a property the instances of which are connected to the instances of ECat by the relation R", where Individual(x), Category(y) and the relation "is-instance-of(x,y)" are one of these triples. We will see how this applies to artifacts and their kinds.

## 14. Intentionally built objects

It is important that for all artifacts x there is exactly one artifact kind K, that x are an instance of.

*Artifact(x): x is an artifact*

*ArtifKind (K): K is an artifact kind*

*is-instance-of(x, K): x is an instance of artifact kind K*

For all artifacts x there exists exactly one artifact kind K that x is an instance of.

(4.) $\forall$ x Artifact(x) $\rightarrow$ $\exists$ !K ArtifKind(K) $\wedge$ is-instance-of(x, K)

In addition to an artifact kind and the concept of that kind, there exist different models of this artifact kind and different designs of these models. Think of the artifact kind "shoe", which could be even classified further in subtypes such as "sneakers", "boots", "sports shoes", "tennis shoes", "sandals" and so on. Still, there are many different models of "tennis shoes" and further, of a certain model, there might be many different designs. Therefore, there shall be introduced the category of a model and a design pertaining to an artifact which are not identical to its kind.

*ArtifDesign(d): d is the design of an artifact x*

*ArtifModel (m): m is the model of an artifact x*

*hasDesign(x,d):Artifact x has design d*

*hasModel(x,m): Artifact x has the model m*

(5.) $\forall$ x Artifact(x) $\rightarrow$ $\exists$ !d ArtifDesign(d) $\wedge$ hasDesign(x,d): Every artifact x has one design d.

(6.) $\forall$ x Artifact(x) $\rightarrow$ $\exists$ !m ArtifModel(m) $\wedge$ hasModel(x,m): Every artifact x has one model m.

Note, that, on the other hand, for an artifact kind K, there exist several models. And for one model, there can be several designs.

(7.) $\forall$ K ArtiKind(K) $\rightarrow$ $\exists$ m ArtifModel(m) : For every artifact kind, there exists at least one model.

(8) $\forall$ m ArtifModel(,) $\rightarrow$ $\exists$ d ArtifDesign(d) : For every artifact model, there exists at least one design.

In order to capture the intention-dependence of artifacts formally, we need a maker S who intends to build an artifact x of a specific kind K with design and model at least implicitly specified. Further, if all artifacts are intentionally built, every artifact has at least one maker, since it is the makers who we ascribe intentions to.

*Maker (s, x): s is a maker of artifact x*

*intend-To-Build (s, x, K): A maker s intends to build an artifact x of a specific kind K.*

For all artifacts x there exists exactly one artifact kind K and at least one maker y and y intends to build artifact x of kind K.

(9.) $\forall$ x Artifact(x) → $\exists$ !K ArtifKind(K) $\wedge$ $\exists$ y Maker(y, x) $\wedge$ intend-To-Build (y, x, K)

A maker can make any object, but they have to intend to make an artifact instance of a specific kind K in order for its result to count as an artifact. Intending to make an object of a certain kind K is sufficient in order to evaluate the result of the actions of making this object according to the description of K and its K-relevant features.

To provide for a more specific relation of building an artifact of a specific kind, model and design, it is possible to further specify the relation "intend-To-Build" into a quintary relation, such that it takes design and model as further arguments.

*intend-To-Build(s, x, m, d, K)***:** *A maker s intends to build an artifact x of model m and design d of a specific kind K.*

## 15. Concept of an artifact kind

The intention to make an artifact of kind K is necessary, but not sufficient to turn an object into the "successful product" of an intention to make an artifact of kind K. In order to succeed to build an artifact of a specific kind K, the maker moreover needs to possess a concept of that artifact kind. A maker m who wants to build an artifact instance x of a particular kind K, then maintains a concept of that artifact kind K. We had specified a maker and the intention to build already as follows.

*Maker (s, x): s is a maker of artifact x*

*intend-To-Build (s, x, K): maker s intends to build an artifact x of a specific kind K*

Now, if the maker s wants to build an artifact instance, they maintain a concept of its artifact kind listing the criterial features of that kind.

*ConceptOf(c, K, Q): c is a concept of artifact kind K encompassing criterial features Q of that kind K*

*has-Concept(s, c, K, Q): a maker s has a concept c of an artifact kind K listing criterial features Q*

For every maker s, for every artifact x, and for every artifact kind K, if s is the maker of x and s intends to build x with artifact kind K, then s has a concept C associated with K and listing its criterial features Q.

(10.) $\forall$ s Maker(s,x) $\forall$ x Artifact(x) $\forall$ K ArtifKind(K) $\wedge$ intend-To-Build (s, x, K)→ $\exists$ c ConceptOf(c, K, Q) $\wedge$ has-Concept(s, C, K, Q)

However, in ascribing a concept to a maker, we allow for his concept of a specific artifact kind K to be wrong, for instance in not being exhaustive with regard to listing the criterial features of an artifact kind or listing wrong features. Note, that the concept a maker has refers to the artifact kind K. This is even the case if a particular instance of an artifact kind has not yet existed. Every new instance of an artifact is of an artifact kind as it can in principle be built again.

As Hilpinen (1993) describes, every artifact kind K has a type-description such that this type- or sortal description lists all the criterial features k1…kn of an artifact kind K that distinguish it from other kinds. I assume that for any artifact kind K, these criterial features k1..kn can be listed.

*ArtifKind (K): K is an artifact kind*

*Is-Criterial-Feature(P, K) : P is a property that is a criterial feature of artifact kind K*

Now, for all artifact kinds there exists at least one criterial feature p.

(11.) $\forall$ K ArtifKind(K) $\rightarrow$ $\exists$ p Is-Criterial-Feature(p,K)

Most of the time, K is defined by exactly one set of criterial feature-properties. One can therefore provide an axiom stating that for all artifact kinds K there exists one set of criterial features Q such that Q defines K.

Criterial-Features(Q,K) : Q is the set of criterial features of artifact kind K

isDefinedby (K, Q): artifact kind K is defined by criterial features Q

(12.) $\forall$ K ArtifKind(K) $\rightarrow$ $\exists$ Q Criterial-Features(Q,K) $\wedge$ isDefinedby (K, Q)

For all artifact kinds K there exists one set of criterial features k1..kn that specify K and differentiate K from other kinds. Of a particular instance x of kind K, it can then be said that x has a criterial feature P of K.

*has-Criterial-Feature(x, P, K): artifact x has criterial feature P of kind K*

## 16. Artifact kinds as interactive kinds

The fact that artifact kinds are specified through criterial features does not exclude that these criterial features cannot change over time. Take the telephone as an example. What we understand as a telephone today includes very different features from what a person in the late 19th century understood as a telephone. Artifact kind descriptions change over time.

Juvshik (2022) rightly points out that artifact kinds are therefore "interactive kinds" (Hauswald, 2016; Khalidi, 2010; Hacking, 1995). The essences of interactive kinds and their criterial features are shaped over time by the audience of their users and makers. Hence, an artifact kind has specifically defined features by and for a specific audience. This audience can be an audience of designers, makers, producers, users, researchers or other stakeholders. Burek (2007) defines similar audiences and processes for function definitions and function ascriptions in GFO. The nature of artifact kinds as interactive kinds calls for audiences that can formulate norms and requirements for these artifact kinds. There can be user requirements, designer requirements, maker/producer requirements or other technical and regulatory requirements. The audience which defines such norms and requirements may vary as well.

## 17. Norms and requirements

It is worth further specifying that it is not sufficient for an agent to have any concept of Ks whatsoever but that they have to have a concept of Ks that more or less conforms with the K-norms, the normative expectations and regulations, of the K-relevant audience A. If the relevant audience of users is only me (private use), I or someone can build an artifact to just my standards. Let us assume I want to build a bike that I can use in my apartment and it is supposed to work just for me. Then I myself define the relevant requirements for that bike to be used within my apartment. Even if I do not explicitly state these requirements somewhere but just implicitly and vaguely assume them when building the bike, these requirements exist. In contrast, if I want to build a bike that can be recognised and used as a bike by anyone in public infrastructures (public use), then I would have to follow not only the standard definition of what it takes to be a bike, but also the safety and other quality standards, requirements and regulations for bikes that allow them to be publicly used on the street and to be functional.

The K-norms are usually defined by an audience that does neither coincide only with the set of producers nor with that of users of an artifact alone. Usually, this audience encompasses standardization bodies, representatives of producers and of users. In cases where the set of producers and users do not coincide, there can also be requirements for the artifact's criterial features and norms by either the producers or the users.

For all artifact kinds K, there exists an audience A that specifies K-norms and K-requirements for that artifact kind K. Further, there can be different audiences specifying different requirements for an artifact kind as well as for a specific artifact model, design and even a specific artifact instance.

*StandardDef (s, K, Q, y): s is a standard definition of an artifact kind K listing criterial features Q and is formulated by audience y*

*ArtifKind (K): K is an artifact kind*

*Audienceof(y): y is a relevant audience of producers, institutions or users*

*specifiesStandDef(y, K, Q, s): audience y specifies a standard definition s with criterial features Q for artifact kind K*

(13.) $\forall$ K ArtifKind(K) $\rightarrow$ $\exists$ s StandardDef (s, K, Q, y) $\land$ $\exists$ Q Criterial-Features(Q,K) $\land$ $\exists$ y Audienceof(y) $\land$ specifiesStand(y, K, Q, s)

So for every artifact kind, there is a standard definition of that kind, there is a set of criterial features of that kind and the standard definition lists criterial features Q and there is an audience y that specifies that standard definition. There does not have to be only one standard definition. Take building standards, there are various different institutional bodies that publish standards and specifications for buildings to the point that there can be even said to exist a market of different competing building specifications and even more so when it comes to sustainability. Standards are however useful because they make people comply with these standards when producing artifacts and this can be important for safety and other reasons such as enabling benchmarking and comparison, enabling a market for services and tools to support compliance with these standards, and in general to develop solutions that scale.

Apart from those standard definitions, there can be requirements of various other forms. The audience to state requirements can be specified into users, researchers, designers, producers, other stakeholders and institutions. Burek (2007) defines similar roles for the process of function design. I will slightly revise his suggestions to make them suitable for the artifact case for the various audiences and complement them by institutions and makers:

*Designer(q,x): q is a designer of artifact x*

*Researcher(q,x): q is a researcher of artifact x*

*User(q,x): q is a user of artifact x*

*OtherStakeholder(q,x): q is a different type of stakeholder with regard to artifact x*

*Maker(s,x): s is a maker of artifact x (as already specified above)*

*Institution(i, K): i is an institution for artifact kind K*

Now, these different audiences can define different requirements with regard to different aspects of an artifact, its kind, its design, its model but also its particularity. I suggest that a requirement n has the form of a proposition p in the following way: A particular artifact x has the property R(x). If relating to an artifact model: A particular artifact model m has the property S(m). Requirements can also be specified for parts of the artifact as well as with regard to the kind, the model, the design and a specific instance of an artifact. The process yields a set of propositions to specify an artifact, its properties and parts, its model, kind, and design. In this way, the set of requirements can eventually be assessed whether it is complete and consistent. I list basic relations for requirements which could be specified further in more extensive work on artifact requirements:

*Krequirement (n, K, y): n is a K-specific requirement for artifact kind K defined by audience y*

*requirement (n, x, y): n is a requirement for the particular artifact instance x specified by audience y*

*Design-requirement (n, d, y): n is a requirement for the design d specified by an audience y*

*Model-requirement(n, m, y): n is a requirement for the model m specified by an audience y*

*Part-requirement(n, p, x, y): n is a requirement for the part p of artifact x specified by an audience y*

*FeatRequirement(n, p, x, y): n is a requirement for the feature p of artifact x specified by an audience y*

*specifies-KRequirement(y, n, K): an audience y specifies requirement n for an artifact kind K*

*specifies-DRequirement(y, n, s): an audience y specifies requirement n for an artifact design d*

*specifies-MRequirement(y, n, m): an audience y specifies requirement n for an artifact model m*

*specifies-Requirement(y,n,x): an audience y specifies requirement n for a particular artifact x*

*specifies-FeatRequirement(y,n,p,x): an audience y specifies requirement n for a feature p of artifact x*

*specifies-Part-requirement(y,n,p,x): an audience y specifies requirement n for a part p of artifact x*

Depending on the situation, there can be different requirements formulated. But for all of these requirements it holds that there is an audience which specifies them.

## 18. Producing an artifact

Apart from knowing what a bike is, and what kind-relevant or criterial features it possesses, an agent in order to successfully build a bike, also has to make the intentional attempt to actually build a bike and has to have the relevant practical knowledge on how to build a bike and bestow the kind-relevant features on an object.

To classify something as an object of an artifact kind K, its maker has to actually make an intentional attempt to build it including the K-relevant features and according to the norms and requirements of an audience, be it successfully or not. Last, artifacts are made out of other stuff or material. Hence, relations that specify the selection and assembling of material into an artifact need to be introduced.

## 19. Artifact production

These conditions all translate into intentional acts of selecting and assembling material and building an artifact of a specific kind K with criterial features k1..kn, complying with the norms and requirements of relevant audiences. I begin with introducing a very general relation to describe the action of intentionally building an artifact instance:

*IntentionalBuild(x, s, K): maker s intentionally builds artifact instance x of kind K*

Note, that I earlier introduced the relation of a maker intending to build an artifact of a specific kind.

*intend-To-Build (s, x, K): maker s intends to build an artifact x of a specific kind K*

These two relations might appear to be similar. They are, however, not identical but only connected. Having an intention or ascribing an intention to someone is different from intentionally carrying out a certain action. The intention can involve carrying out many different actions. Here, I am referring to a distinction Michael Bratman (1987) draws in his essay on "Two Faces of Intention": "In acting intentionally there is something I intend to do; but this need not be what I do intentionally." To produce an artifact of a certain kind is something I intend and have an intention for. But the various intentional actions that carrying out this intention involves, differ and may even be carried out by different actors. But those actions are still what these actors do intentionally. Hence, as Bratman notes, there are actions that are "performed intentionally in the course of executing a certain intention" (1987:119).

I suggest further specifying these intentional production actions of a maker as specifications or subclasses of the act of intentionally building an artifact. For instance, a maker intentionally bestows features on an object.

*IntentionalBestow (x, s,  K, Q) : a maker s intentionally bestows K-criterial features Q on an object x of kind K*

Further, artifacts are always made out of something. In order to capture this process of modification, assemblage or appropriation of other objects into artifacts of a specific kind, a maker has to select the adequate material and assemble them in the correct way to build an instance of an artifact kind K that they want to build. Here, I suggest dividing production actions into a selection process of selecting adequate material and assembling the material. There could possibly be added actions of shaping or structuring or composing the material.

*IntentionalSelect(s, m, x, K): a maker s intentionally selects material m for artifact x of kind K*

The list provided here is not exhaustive of all possible actions involved in a production process and could be elaborated further. Still, I would like to point to the fact that there can be many makers involved in these intentional actions. Especially in cases of industrial production, the different steps of designing, planning, modeling, testing, assembling, prototyping and evaluating artifacts are assembled around multi-actor value and supply chains. These cases of collaborative or industrial production of artifacts seem to require a detailed planning theory of intentions as outlined by authors such as Bratman (1987), Robert Audi (1991), and Per Galle (1999). The actions of defining, designing, modeling, testing, prototyping, assembling and evaluating an artifact, happen all in the course of executing the intention to build an instance, or many instances, of that artifact kind.

## 20. Intentional Object and Design

The design, modeling and planning processes of artifacts of a certain artifact kind have not been discussed much yet in the field of formal ontology of artifacts. Activities of planning, designing, or modeling, including the formulation of requirements for an artifact, imply that an agent can anticipate the planned object in descriptions of various kinds—whether linguistic, mathematical-logical, numerical, or imaginative. In such contexts, mental representation as well as an object model can serve for the gradual specification and characterization of an object that is to be present in a complete form—digitally or materially - thereafter. Complete form here refers to the properties and capabilities planned by an actor or a group of actors, potentially including the abilities to perform specific functions. The intended object or the initially mentally or otherwise represented object prescribes a normative type, based on which the fully realized object should be evaluable. The type defined by descriptions—the properties defined for the intended object such as size, shape, color, material, capabilities—plays a role in the evaluation and testing phase of the object. Assuming the requirement for the object is that it can rotate under certain conditions, this prior description of the ability and the conditions under which the object should be able to perform the ability serves as a normative stipulation for the subsequent evaluation and testing of the object: Can the object truly rotate under certain conditions? If it meets the formulated requirements, its realization is successful.

The following specifications introduce descriptions of a mental representation of an actor of an artifact they would like to build. Every artifact is of a certain kind and has a design and a model. Accordingly, a maker of an artifact can have, however vague, a certain mental representation of the object they want to create. I do not contend that this must necessarily be the case, but in professionalized contexts of artifact production, it certainly is. To facilitate the description of these, I introduce the following relations.

*MentRepresArt(p, x, K, m, d, Q): p is a mental representation of an artifact x of kind K and model m, design d with features Q*

*hasMentRepresentation(s, p, x) : a maker s has a mental representation p of an artifact x*

Further, the maker can transform this mental representation into an actual design and model for an artifact.

*transforms_to(s, p, x, DesAct, ModAct): a maker s transforms a mental representation p of artifact x into an actual design d and model m*

The process of transforming virtual, imagined features into actual design, modeling or a prototype allows for the iteration of this process to successively enrich the design and model of an artifact. These relations at least provide a beginning of what is needed to fully capture these processes in formal terms. Further research needs to be done in order to enrich this discussion towards a more precise account of artifact representation and design.

## 21. Conclusion

This paper aimed to bridge a gap within the top-level ontology GFO, advancing it towards integrating descriptions for artifacts—material objects intentionally constructed to fulfill specific purposes or roles. Leveraging the foundation of its space module (Loebe et al., 2021), where material objects are defined in relation to space, and considering work by Burek (2007, 2021) on functions and foundational work for GFO by Herre (2013, 2019), basic categories for artifacts, their kinds and their makers and users, were developed. For the phase of planning and designing an artifact of a specific kind, relations of formulating requirements were provided and foundational production actions were defined. In order to describe how intentions guide planning, designing, and production processes over time, the paper provided relations to capture the mental presentation of an artifact object and its translation into a materialized form. These categories serve as a foundation for further development in this domain.

*In addition to this paper, I have compiled a RDF (Resource Description Framework) Turtle description for the artifact axioms developed within it. This description is hosted in the GitHub repository under the URL: https://github.com/hannafiegen/ArtifactGFO. The RDF Turtle description is available under the Creative Commons Attribution 4.0 International License (CC BY 4.0), promoting open access and reuse of these contributions.*